\documentstyle[psfig]{caosp}

\begin{document}
%
\hauthor{G. Scholz {\it et al.}}
\title{Magnetic field distribution and element concentration on the CP2 star 
CU~Virginis}
\author{Yu.V. Glagolevskij
\inst{1}
\and E. Gerth
\inst{2}
\and G. Hildebrandt
\inst{2}
\and G. Scholz
\inst{2}}
\institute{Special Astrophysical Observatory of Russian AS, Nizhnij Arkhyz 
357147, Russia\\
Internet: glagol@sao.stavropol.su
\and Astrophysikalisches Institut Potsdam, Telegrafenberg A27, D-14473
Potsdam, Germany\\
Internet: egerth@aip.de, ghildebrandt@aip.de, gscholz@aip.de}

\date{December 15, 1997}

\maketitle

\begin{abstract}
We search for a relation between the published distributions of different 
elements and the calculated magnetic field structure, following from a 
dipole-quadrupole configuration, of the CP2 star CU Vir. The highest 
concentration of individual chemical elements on the stellar surface 
coincides obviously with the regions of the highest values of the magnetic 
field strength.

\keywords{stars: chemically peculiar stars - magnetic fields - element
abundances}

\end{abstract}

\section{Introduction} 
The B9pSi star CU Vir (HD 124224, $P_{\rm rot}$~=~0\fd52, e.g. 
Weiss et al. 1976), has an amplitude of the effective magnetic field 
curve of about 1000 gauss and shows very large spectral variations of 
helium and silicon. Relations between the structure of the magnetic field 
and the distribution of some chemical elements on the stellar surface should 
exist according to, e.g., Michaud (1970), Glagolevskij (1994), and 
Hatzes (1997). For our investigation we use the values of the longitudinal 
magnetic field, $B_{\rm eff}$, measured photoelectrically by Borra and 
Landstreet (1980) and the distribution of He, Si, and Mg over the stellar 
surface derived by Goncharskij et al. (1983), Hiesberger et al. (1995), 
Kuschnig et al. (1997), and Hatzes (1997).

\section{Magnetic model of CU Vir} 
The magnetic field structure has been calculated on the assumption 
that the variation of the $B_{\rm eff}$ values observed is caused by a 
dipole-quadrupole configuration of the magnetic field and the model 
of the oblique rotator. Fig. 1 represents the observations (dots) of CU Vir 
versus the rotational phase. The points differ distinctly from a sinusoidal 
curve, which would exist for a dipole field.\\ 
\begin{figure}
\psfig{figure=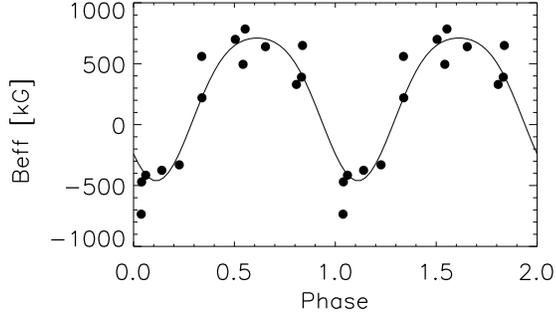,height=5cm}
\caption{Variation of the longitudinal magnetic field after 
Borra~\&~Landstreet (dots) and our model (curve)}
\end{figure}
For the magnetic field modelling we take the angle $i$~=~60\degr~between the 
rotation axis and the line of sight and we set the angle $\beta$ = 65\degr, 
describing the inclination between the rotation axis and the dipole axis. 
The magnetic sources are located at $R_{\rm *}$~=~0.1 from the centre and 
produce at the stellar surface the magnetic field strengths of the dipole 
of +1.5 and $-$1.5 kG and of the quadrupole of +3.1, $-$3.1, +3.1, and $-$3.1~
kG. The calculated variation of $B_{\rm eff}$~is represented in Fig. 1.
 
\section{Magnetic structure and surface maps}
\subsection{He} 
We have taken the distribution of He over the stellar surface from Hiesberger 
et al. (1995) and Kuschnig et al. (1997). The He spots obviously coincide
with the regions of the maximal field strength. Differences between the 
distribution maps quoted above occur especially in the region of the broad 
positive magnetic maximum. The solution of Hiesberger et al. appears to be 
closer to the magnetic field structure derived from our model. In the 
vicinity of the maximal negative pole only a weak He spot exists. An 
explanation could be that we observe only the edge of a large spot, but this 
spot is practically invisible because of the large inclination of the star.

\subsection{Si, Mg}
To search for a relation between the magnetic field and the element Si we 
use the Si map distributions of Goncharskij et al. (1983), Kuschnig et al. 
(1997), and Hatzes (1997). Only a general tendency can be seen: the Si 
abundance is lower in the region of the broad positive maximum, whereas in 
the region of the negative magnetic field extremum the abundance seems to 
be higher.\\ 
Al\'ecian and Vauclair (1981) and M\'egessier (1984) have shown the 
importance of the horizontal component of the diffusion velocity in magnetic  
Ap stars. They postulated an overabundance of Si in these regions. 
To test these suggestions we calculated the isolines of the horizontal 
magnetic field lines from our dipole-quadrupole model. Comparing 
with Hatzes Si map a concentration of this element in the areas with the 
horizontal direction of the magnetic field lines seems possible.\\ 
The behaviour of Mg, of which only the distribution map of Goncharskij et al.
is known, is obviously quite similar to that of Si.

\section{Discussion}
Our dipole-quadrupole model yields two maxima at the positive halfwave of 
the magnetic field and a very satisfactory agreement between the calculated and
the observed behaviour of $B_{\rm eff}$. We are not able to fit the strong 
anharmonic variation of $B_{\rm eff}$~by a decentred dipole field. The two 
observed He spots seem to support the assumption of a dipole-quadrupole 
configuration.\\ 
The maps proposed for Si allow to explain the element distribution over the 
surface by a concentration of the element at the magnetic poles as well
as in the regions where the magnetic field lines have a horizontal 
direction. To overcome the existing discrepancies, more accurate observations 
are necessary.\\ 

\acknowledgements
The authors thank Dr. T.A. Ryabchikova 
for sending us unpublished results from the paper of Kuschnig et al. (1997).

{}

\end{document}